\documentclass[%
 reprint,
 amsmath,amssymb,
 aps,
]{revtex4-2}
\pdfoutput=1
\usepackage{array}
\usepackage{graphicx}% Include figure files
\usepackage{dcolumn}% Align table columns on decimal point
\usepackage{bm}% bold math
\usepackage{upgreek}
\begin{document}

\preprint{APS/123-QED}

\title{Simultaneous multi-band radio-frequency detection using high-orbital-angular-momentum states in a Rydberg-atom receiver}% Force line breaks with \\

\author{Gianluca Allinson}
 \email{gianluca.allinson@durham.ac.uk}%Lines break automatically or can be forced with \\
\author{Matthew J. Jamieson}
\author{Andrew R. Mackellar}
\author{Lucy Downes}
\author{C. Stuart Adams}
\author{Kevin J. Weatherill}
\affiliation{Department of Physics, Durham University, South Road, Durham, DH1 3LE}

\date{\today}% It is always \today, today,

\begin{abstract}
We demonstrate simultaneous detection of radio-frequency (RF) fields ranging from the very high-frequency (VHF) band (128 MHz) to terahertz frequencies (0.61 THz) using a caesium Rydberg-atom receiver. The RF fields are concurrently applied to a series of atomic transitions involving states of increasing orbital angular momentum, where the energy separations become progressively smaller, allowing access to a very wide range of radio frequencies. We show that the optical response of the system in the presence of the RF fields can be reproduced theoretically using a simple Lindblad-master-equation approach. Furthermore, we demonstrate experimentally that a series of amplitude-modulated tones can be detected simultaneously using multiple carrier frequencies. This demonstration opens the way for RF communications across multiple bands simultaneously using a single optical receiver. 
\end{abstract}

\maketitle

%\tableofcontents

\section{\label{sec:level1} Introduction}
Rydberg atoms, i.e., atoms in which an electron has been promoted to a highly excited state with large principal quantum number $n$, are increasingly finding use in technological applications~\cite{Adams2020}. The polarizability of atoms, and the transition strength between neighboring electronic states, both increase with increasing $n$, making Rydberg atoms highly sensitive to both AC and DC electric fields. 
This sensitivity has been exploited to demonstrate atom-based radio-frequency (RF) electric field metrology~\cite{Sedlacek2012}, RF sensing~\cite{superhet}, THz imaging~\cite{Wade2017, Downes2020} and RF communications~\cite{Song:19}.
Most Rydberg-atom-based RF sensors make use of electromagnetically induced transparency (EIT)~\cite{Mohapatra2007}, a coherent and non-destructive process that couples the properties of the Rydberg state to an optical transition from the atomic ground state, thereby allowing the perturbative effect of an incoming RF electric field on the Rydberg state to be mapped onto an optical probe field, which is detected using a photodetector~\cite{Mohapatra2008,Tanasittikosol_2011}. This EIT technique effectively enables extremely sensitive and SI-traceable measurements of RF electric fields to be extracted from an optical signal~\cite{shaffer-tutorial}. The last decade has seen an explosion of development in this research area, with Rydberg-atom receivers demonstrating high-performance in the measurement of RF field properties, from precision amplitude~\cite{superhet}, frequency~\cite{gordon2019} and phase~\cite{superhet, simons2019} measurements to polarization~\cite{Sedlacek2013}, angle of arrival~\cite{Robinson2021} and simultaneous detection of multiple frequencies~\cite{meyer2023simultaneous, jayaseelan2023eit}.\par
One of the major strengths of a Rydberg-atom RF receiver, when compared to other RF technologies, is its potential to detect an enormous frequency range using a single device~\cite{meyer2020assessment}. Indeed, Rydberg-atom receivers have demonstrated RF field detection at frequencies ranging from below 1~kHz~\cite{Yuan2020} to above 1~THz~\cite{Chen:22}. In the low frequency, or quasi-DC detection regime~\cite{Mohapatra2008,meyer2020assessment,meyer2021waveguide}, Rydberg receivers can achieve continuous frequency coverage but with limited sensitivity. Whereas, in the higher-frequency, AC detection regime, state-of-the-art sensitivities can be achieved~\cite{Sedlacek2012, superhet} but only over a small range of frequencies around the discrete values defined by resonant atomic transitions. Furthermore, the resonant frequency of Rydberg-Rydberg transitions decreases with increasing $n$ meaning that states with very high $n$ ($>$ 100) are typically required to access frequencies below 1~GHz in the sensitive AC regime. Gaining access to low frequencies using higher $n$ states places demanding requirements on the laser power needed for experiments. Consequently, recent work has focused on gaining access to lower frequency bands via the use of higher angular momentum states~\cite{brown2023very,elgee2023satellite}. This previously involved the use of a three step laser excitation scheme~\cite{carr2012three} plus an RF field to access transitions of the form $n$F$\to n^{\prime}$G. The use of higher angular momentum measurements have extended the range of resonant RF transitions to $\sim$1~GHz and below at more accessible principal quantum numbers ($n<$ 70).\par
In this work, we take the principle of using higher-angular-momentum states much further, to demonstrate that a Rydberg atom ensemble is capable of simultaneously receiving multiple RF signals (seven in this case). They are a set of discrete carriers which range from the very high frequency (VHF) band (30 - 300~MHz) to the THz band (0.3 - 3~THz), ultimately spanning 12 octaves. Over said RF signals, amplitude modulated tones are transmitted and detected simultaneously by a mapping the modulation to the optical probe. This is achieved by using two optical photons to reach a Rydberg state and a sequence of cascading RF fields that resonantly couple higher-angular-momentum states in the atom. Furthermore, we demonstrate that the VHF range can be accessed at significantly lower principal quantum numbers, $n=17$, where only modest laser power is required. 
\section{Method}
\begin{figure}[h]
\centering\hspace*{-0.2cm}
\includegraphics[width=1\linewidth]{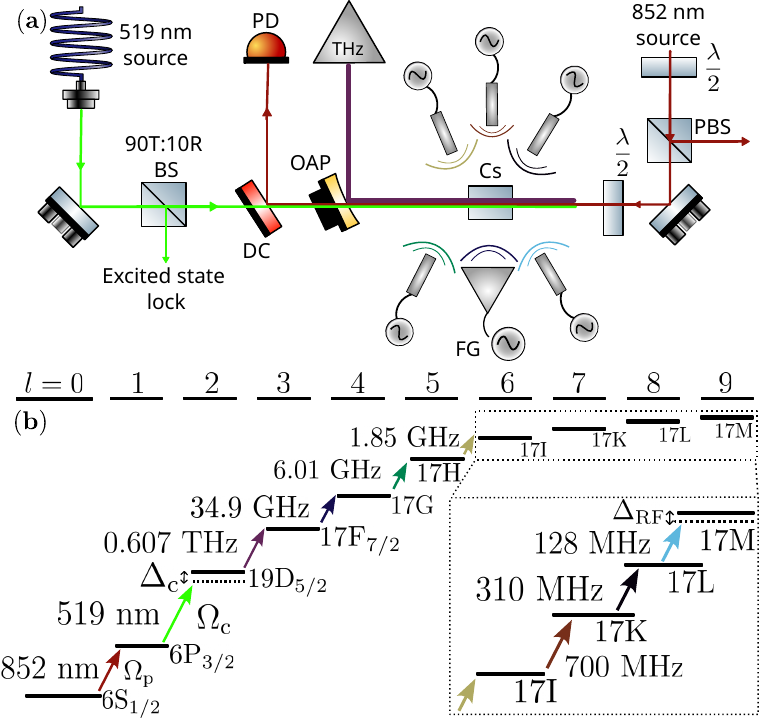}
\caption{(\textbf{a}) Experimental layout of the Rydberg receiver with 852~nm probe, 519~nm coupling, and 0.607~THz fields overlapping within the Cs vapor cell. DC - dichroic mirror, OAP - off-axis parabolic mirror, (P)BS, (polarizing) beam-splitter, FG - frequency generator, PD - photodiode, $\lambda$/2 - half-waveplate. A pyramidal RF gain horn and five whip antennas provide the RF fields.  (\textbf{b}) An energy-level diagram showing the electronic states used. A zoomed inset shows the 17$\ell$ states with $\ell \geq 6$ for clarity. Transitions are color coded in (\textbf{a}) to their counterpart in (\textbf{b}) and subsequent figures.}
\label{fig:1}
\end{figure}
Simultaneous multi-band RF detection is implemented using laser spectroscopy of an atomic vapor.  
A schematic diagram of the experimental arrangement is shown in Fig~\ref{fig:1} (a). A linearly polarized 852~nm probe beam (solid red line), resonant with the $6$S$_{1/2}, F=4 \rightarrow 6$P$_{3/2}, F'=5$ transition, was passed through a 1~cm long cuboidal cuvette containing caesium (Cs) vapor at room temperature (23 $^\circ$C) and then detected using a photodiode (PD). A counter-propagating and linearly polarized 519~nm coupling laser (solid green line) with detuning $\Delta_{\rm c}$ from the $6$P$_{3/2}\rightarrow19$D$_{5/2}$ transition was overlapped with the probe field using a dichroic mirror (DC). The probe and coupling beams have powers of 20 $\upmu$W and 21~mW and 1/$e^{2}$ beam waists of 1.20(7)~mm and 0.96(7)~mm respectively. The probe beam is derived from a \textit{Toptica DLpro} laser and stabilized using polarization spectroscopy~\cite{pearman2002polarization}. The coupling beam is derived from a home-built second-harmonic generation system of a 1030~nm ECDL based on the design of Legaie {\it et al.}~\cite{legaie_sub-kilohertz_2018}. Cavity power build-up is stabilized by a modified Hänsch-Couillaud lock~\cite{hansch1980laser, vainio2011cavity}. The coupling laser is frequency stabilized by an excited-state lock using nonlinear polarization spectroscopy~\cite{meyer2017nonlinear}. 

\begin{figure*}[ht!]
\centering\hspace*{-0.6cm}
\includegraphics[width=0.9\linewidth]{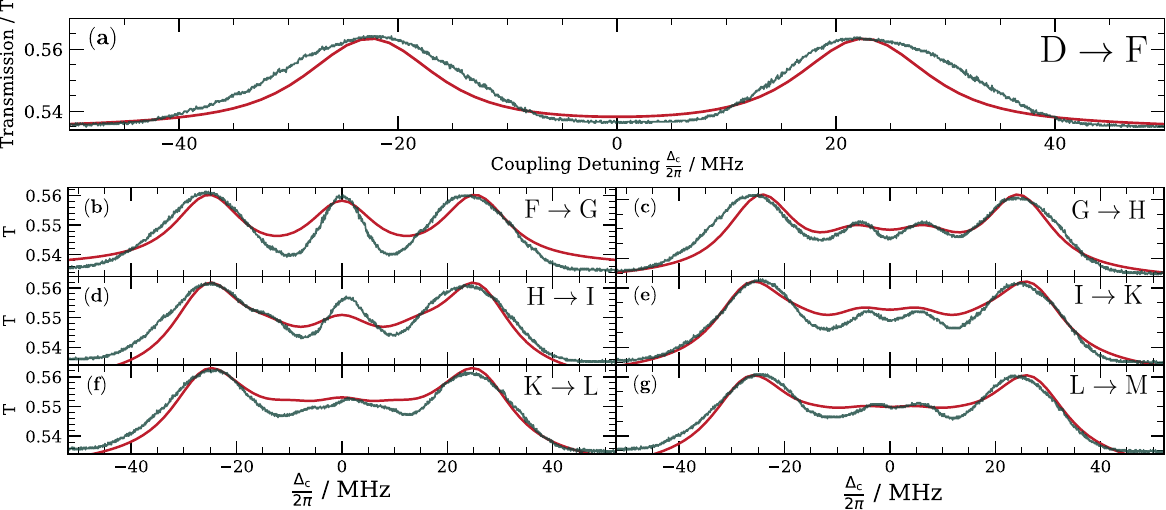}
\caption{Absolute probe transmission traces from the experiment (green) with each subsequent addition of a field in the ladder system overlaid with an n-level master equation model (red). The frequency scale is calibrated via the 6P$_{3/2}$ $F$ = 4, $F$ = 5 separation. (\textbf{a}) The probe and coupling field on with the THz field resonant to the 19D${}_{5/2}\to$17F${}_{7/2}$ transition. The subsequent microwave fields are then resonant with the (\textbf{b}) 17F${}_{7/2}\to$17G (\textbf{c}) 17G$\to$17H (\textbf{d}) 17H$\to$17I (\textbf{e}) 17I$\to$17K (\textbf{f}) 17K$\to$17L (\textbf{g}) and 17L$\to$17M transitions. Each transition is labelled with the respective $\ell \to {\ell}^{\prime}$ for clarity.}
\label{fig:2}
\end{figure*}

A linearly polarized terahertz (THz) beam (solid purple line) resonant with the $19$D$_{5/2}\rightarrow17$F$_{7/2}$ transition at 0.607~THz is launched from a diagonal horn antenna, collimated using a polytetrafluoroethylene (PTFE) lens (not shown) and focused in the Cs cell using an off-axis parabolic (OAP) mirror. The OAP contains a 2~mm through-hole to allow the laser beams to copropagate with the THz beam. The THz beam is derived from a \textit{Virginia Diodes} amplifier multiplier chain (AMC) and is estimated to have a power of 10~$\upmu$W and beam waist of approximately 1.5~mm within the cell. The transmission of the probe beam is recorded as the frequency of the coupling beam ($\Delta_{\rm c}$) is varied. The EIT spectrum obtained in the presence of the THz field is presented in Fig~\ref{fig:2} (a). 
The RF fields are generated using frequency generators and emitted via commercial Wi-Fi, cellular, and radio $\lambda/4$ monopole whip antennas that are placed approximately 10-20~cm from the cell where there is clear line-of-sight. The inclusion of each additional field changes the transmission profile of the probe beam as shown in Fig~\ref{fig:2} (b) - (g).
\begin{table}[h]
    \centering
    \begin{tabular}{ccccc}
        \hline
          \multicolumn{2}{c}{Atomic transition}& Final $\ell$ & RF (GHz)& DME ($ea_{0}$)  \\
         \hline \hline
           \multicolumn{2}{c}{$19$D$_{5/2}$ $\to$ $17$F$_{7/2}$}& 3 & 607& 354  \\
           \multicolumn{2}{c}{$17$F$_{7/2}$ $\to$ $17$G$_{9/2}$}& 4 &34.9& 423  \\
           \multicolumn{2}{c}{$17$G$_{9/2}$ $\to$ $17$H$_{11/2}$}& 5 &6.01& 414  \\
           \multicolumn{2}{c}{$17$H$_{11/2}$ $\to$ $17$I$_{13/2}$}& 6& 1.85&  406  \\
           \multicolumn{2}{c}{$17$I$_{13/2}$ $\to$ $17$K$_{15/2}$}&7  &0.700& 395  \\
           \multicolumn{2}{c}{$17$K$_{15/2}$ $\to$ $17$L$_{17/2}$}& 8 &0.310&  383  \\
           \multicolumn{2}{c}{$17$L$_{17/2}$ $\to$ $17$M$_{19/2}$}&  9&0.128&  368  \\
           \hline
    \end{tabular}
    \caption{A list of the electronic states that are coupled by the series of cascading RF fields. DME - (Radial) dipole matrix element. Whilst the fine structure splitting is not resolved beyond F, we label the strongest coupled $j$ state for completeness. Note that by convention `J' is omitted as the symbol for the $\ell$ = 7 azimuthal quantum number.}
    \label{tab:1}
\end{table} 
The transitions used are shown schematically in Fig \ref{fig:1} (b) with their frequencies and numerically calculated dipole matrix elements (DMEs) presented in Table~\ref{tab:1}~\cite{vsibalic2017arc}. We exploit the fact that beyond $\ell\geq 3$, transitions of the form $n\ell \to n(\ell +1)$ allow for lower frequency bands to be addressed.
The rapidly decreasing overlap of the electron wavefunction with the atomic core results in extremely small quantum defects, and hence small energy separations, of high-$\ell$ states~\cite{gallagher1994rydberg}. Each RF field is applied such that it is resonant with the transition to the next highest angular momentum atomic state. A higher $\ell$ state can only be coupled if the lower $\ell$ state is coupled by the relevant RF field. Consequently, the 17L$\to$17M coupling at 128~MHz is dependent on all previous states to be coupled by their resonant RF and optical fields.
\par
When the high $\ell$ states are all simultaneously coupled by their respective resonant RF fields, we can amplitude modulate the fields and detect the baseband tones on the optical signal. For all data, readout is taken from the photodiode via USB-DAQ of an oscilloscope and no lock-in detection/amplification is used. 
\section{Results}
\subsection{Multi-level EIT Cascade Scheme}
The experimental spectra in Fig~\ref{fig:2} can be understood in terms of a $N$-level EIT cascade scheme~\cite{mcgloin2001electromagnetically}. With the probe laser locked, a transparency window in the probe absorption profile is expected around $\Delta_{\rm c}$ = 0, resulting in a coherent transmission feature as the coupling laser is detuned across this resonance. This phenomenon is known as electromagnetically induced transparency (EIT)~\cite{mohapatra2007coherent}. Applying the THz field results in a reduction in transmission of the probe field at $\Delta_{\rm c}$ = 0. For large THz field strengths, this results in the Autler-Townes (AT) splitting seen prominently in Fig~\ref{fig:2} (a). With the addition of another field, there is again transparency in the line center. This is apparent in the data in the left-hand column of Fig \ref{fig:2}, showing the distinct increase in the probe transmission at zero detuning of the coupling field for an odd number of levels. For an even number of levels, there is absorption about the line center, shown in the right hand column of Fig~\ref{fig:2}. When the subsequent RF fields are applied, the previous resultant AT splitting is further shifted from the line center and decreases in amplitude. These features become hard to distinguish in the lineshape, nevertheless modulation of these fields causes a change to the transmission of the probe beam.
The atomic ensemble is modelled using a Lindblad master equation approach, which gives a time evolution of the density matrix, $\hat{\rho}$:
\begin{equation}
    \frac{d\hat{\rho}}{dt} = -\frac{i}{\hbar}[\hat{H},\hat{\rho}] + \hat{L}_{\textrm{atom}} + \hat{L}_{\textrm{dephasing}},
\end{equation}
where $\hat{H}$ is the Hamiltonian of the atom-light system, $\hat{L}_{\textrm{atom}}$ is the Linblad superoperator describing spontaneous decay, and $\hat{L}_{\textrm{dephasing}}$ is dephasing due to the laser linewidths or collisional effects. The series of differential equations formed are then solved in the steady state to give the susceptibility of the atomic ensemble~\cite{downes2023simple}. 
The system is repeatedly solved for varying velocity classes and integrated over a bounded velocity distribution to obtain the Doppler broadened profile. 
Outside the weak probe regime, numerical solutions are needed. Such a computation takes a significant amount of time on modern computer systems, leaving computer minimization for such a problem impractical. Model inputs, such as the Rabi frequencies for each field, were varied around the experimental values to give the best fit by eye. \par
The resulting model spectra are shown by the red lines in Fig~\ref{fig:2}, along with those from the experiment. Only one $m_{j}$ state is used for each level; including all sub-levels would result in a model consisting of 110 levels. Fine structure beyond the F state is ignored but note that the $j = \ell +\frac{1}{2}$ states are coupled the strongest and are used in the calculation of dipole matrix elements for G, H and beyond. 
We find good agreement with the model in the generality of adding additional levels. The parameters used for the model were $\Omega_{\rm p} = 2\pi \times 1.8$~MHz, $\Omega_{\rm c} = 2\pi \times 3.7$~MHz and $\Omega_{\rm THz} = 2\pi \times 46$~MHz. The remaining Rabi frequencies of the RF fields varied by a few MHz but were approximately $2\pi \times 20$~MHz.

\begin{figure}
\centering\includegraphics[width=\linewidth]{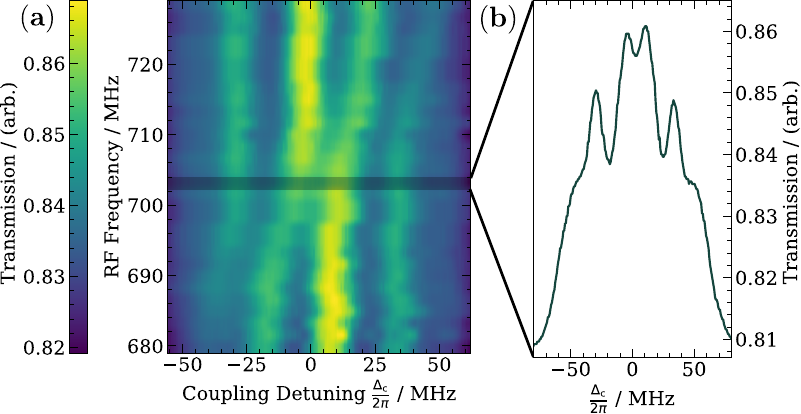}
\caption{Spectral map of the I $\to$ K transition as the RF field is swept and coupling laser is detuned across resonance. (\textbf{a}) A color map made up of the resultant probe transmission traces in response to scanning the RF field over resonance for the 17I$\to$17K transition at $\sim$700 MHz. (\textbf{b}) A plot of the probe transmission against the coupling detuning for an RF field of 0.703 GHz - a grey bar shows the EIT trace which makes up that respective part of the color map.} \label{fig:3}
\end{figure}

Fig \ref{fig:3} (a) shows behaviour of the lineshape in response to detuning the RF field from resonance. Here, the probe laser power ($\sim$100 $\upmu$W) and the power of the RF fields are increased in order to improve signal to noise and visibility of the distinct AT splittings arising from each of the RF field couplings. The contrast is seen in Fig  \ref{fig:3} (b) when compared to Fig \ref{fig:2} (e) whereby each of the distinct AT peaks are separated and can be distinguished.

The transmission window in the lineshape persists over a large detuning range, $\Delta_{\rm RF} = \pm 20$~MHz. As this is an even level system, the induced transparency from the presence of the previous field slowly shifts about the line center and splits until symmetric at $\Delta_{\rm RF} = 0$~MHz. Because of the high field strengths used, the resonance of the feature is seen at $\sim$705~MHz compared to the previously observed 700~MHz used in Fig~\ref{fig:2} (e). This can be attributed to the large AC Stark shifts that culminate for each RF field that is on resonance. 
\subsection{Amplitude Modulation of Resonant RF Fields}
The RF fields were modulated to demonstrate simultaneous detection across the various frequencies that are resonant with the atomic transitions. 
A change in each RF field amplitude will independently change the transmission of the probe beam. When amplitude modulated, the baseband tones on each RF field can then be extracted via performing a fast Fourier transform (FFT) of the probe transmission signal. In a Rydberg atom detector, demodulation of the RF frequencies are performed by the atom in and of itself, so no additional demodulation of the signal is necessary.

\begin{figure*}[ht!]
\centering\hspace*{+0.3cm}
\centering\includegraphics[width=1\linewidth]{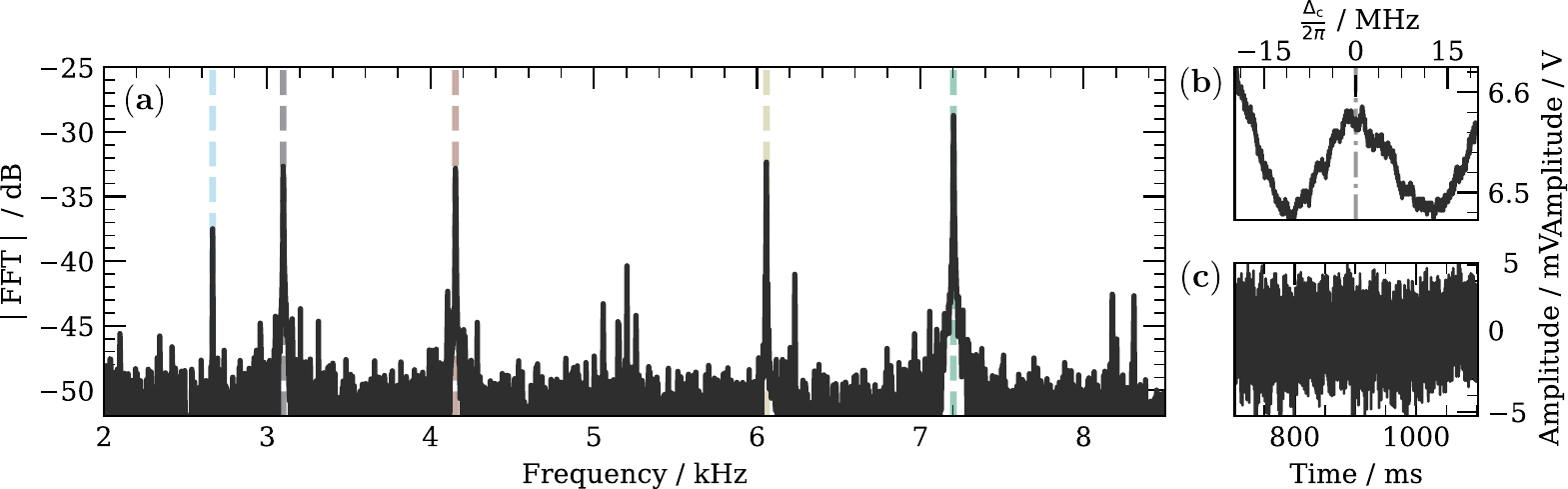}
\caption{Simultaneous detection of five independent AM tones using five of the seven RF carriers.  (\textbf{a}) The modulus of an FFT of the probe transmission at $\Delta_{\rm c} = 0$. The amplitude-modulated tones used were 7.2 kHz, 6.05 kHz, 4.15 kHz, 3.1 kHz, and 2.66 kHz on carrier frequencies of 6.01 GHz, 1.85 GHz, 700 MHz, 310 MHz, and 128 MHz, respectively. A dashed line is shown on each of the peaks for visual aid. (\textbf{b}) An inset showing a typical scan of an EIT feature during modulation of the RF fields. A vertical dashed line shows the point at which the coupling laser is locked. (\textbf{c}) A typical trace before it is discrete Fourier transformed.}
\label{fig:4}
\end{figure*}

Each electronic state detailed above were simultaneously coupled by the probe, coupling, THz and the 6 RF fields with powers similar to those used in the spectra of Fig.~\ref{fig:3}. A baseband tone in the kHz frequency range was modulated on each carrier RF field and detailed in Fig.~\ref{fig:4}. The modulation index of the AM used on all of the carrier frequencies was 0.5. The resulting modulus of the fast Fourier transform of the probe signal is shown in Fig~\ref{fig:4} (a).

The data show that the baseband tone from each of the independent carriers can be recovered from the probe signal. The decreasing strength of the FFT signal as the carrier frequency decreases can somewhat be attributed to the decreasing DME of the transitions. 
In order to remove servo noise that we observed in the FFT, the data presented is taken by disengaging the coupling laser frequency lock and subsequently saving a trace.  Due to technical limitations, the THz source and 34.9~GHz MW generator could not be amplitude modulated. However, there is nothing to suggest that amplitude modulation of these transitions could not be detected.
\section{Discussion}
In principle, the number of fields that could be applied is limited by the possible angular momentum states, ($n-1$). 
At higher principal quantum number, the sensitivity to the fields would increase, and the splitting between neighboring $\ell$ states would decrease. For example, neighboring angular momentum transitions $\ell\geq8$ at higher $n$ (40 -70) would correspond to resonant fields in the medium frequency (300~kHz - 3~MHz) or high frequency (3~MHz - 30~MHz) radio bands. 
\par
We note that at low $n$, many perturbing effects such as DC Stark induced state mixing and Rydberg-Rydberg collisions are significantly reduced~\cite{gallagher1994rydberg}. Despite the lower sensitivity to external RF fields, electrometry at low principal quantum number may be more suitable for calibration of an RF field or metrology applications. 
A sensitivity assessment of the method to each of the applied RF fields is beyond the scope of this paper as the sensitivity for each transition depends upon the driving strength of all previous transitions and  the parameter space is very large.  We note that the experimental requirements used here are relatively modest, as we use low coupling-laser power and do not rely on lock-in detection.
The use of other well-characterized antennas for the lower frequency fields would allow a better estimation of the electric field strengths involved. The $\lambda/4$ monopole antennas used are not well-characterized, are omnidirectional, and have poor efficiency, resulting in significant power loss before the cell. 
\par
The method presented above could be used to investigate higher-angular-momentum states which are not as well studied as the S, P, D, and F states. There is currently no published work on quantum defects in Cs higher than G$_{7/2}$ known to the authors~\cite{weber1987accurate,lorenzen1984precise}, whilst some exists for rubidium~\cite{Civiš_2012,berl2020core,PhysRevA.102.062817}. Transitions were found in this work by sweeping the frequency of the RF fields to observe the largest change in transmission of the probe beam at low RF power. Initial estimations of the energy levels were given by \verb|arc|, which uses an approximation to estimate $\delta_{\ell}$ for $\ell\geq4$ based on known quantum defects~\cite{robertson2021arc}. The observed resonances were found to be in general agreement and were additionally compared to the scarce available literature~\cite{safinya1980resonance}. Discrepancies on the order of several MHz are found for $\ell \geq 7$, which could be attributed to the culminating AC Stark shift of the many resonant fields. There is no control of stray magnetic/electric fields which may cause further splitting of the numerous $m_{J}$ states.

\section{Conclusion}
Simultaneous detection of RF fields across 12 octaves by an optically coherent method is demonstrated. The method opens up a new way of conducting radio frequency electrometry using a cascade of RF fields across adjacent angular momentum coupled states. Further work should be done to characterize and optimize the technique both for higher principal quantum number and 3-photon electrometry methods.

The data presented in this paper are available at~\cite{data}.

\begin{acknowledgments}
The authors would like to thank Robert Potvliege and Popoola Waisu for useful discussions, Ifan Hughes for careful reading of the manuscript,  Jonathan Pritchard for SHG laser designs and Mike Tarbutt for the loan of equipment. \par
We acknowledge the UK Engineering and Physical Sciences Research Council under grants EP/R002061, EP/V030280, EP/W009404 and EP/W033054.

\end{acknowledgments}

\bibliography{apssamp}% Produces the bibliography via BibTeX.

\end{document}